# Blockchain and Sustainability: A Tertiary Study

Shanshan Jiang
SINTEF Digital
Trondheim, Norway
shanshan.jiang@sintef.no

Kine Jakobsen
SINTEF Nord
Tromsø, Norway
kine.jakobsen@sintef.no

Letizia Jaccheri and Jingyue Li
Norwegian University of Science and Technology
Trondheim, Norway
{letizia.jaccheri, jingyue.li}@ntnu.no

*Abstract*—Blockchain is an emerging technology with potential to address issues related to sustainability. Literature reviews on blockchain and sustainability exist, but there is a need to consolidate existing results, in particular, in terms of Sustainable Development Goals (SDG). This extended abstract presents an ongoing tertiary study based on existing literature reviews to investigate the relationship between blockchain and sustainability in terms of SDGs. Results from a pilot analysis of 18 reviews using thematic analysis are presented.

## I. THE CONTEXT

The topic of sustainability is receiving increasing attention and commitments worldwide. The United Nations' 2030 agenda for sustainable development has identified 17 Sustainable Development Goals (SDGs) with 169 targets [1]. A widely accepted concept is Elkington's triple-bottom-line concept, which assesses sustainability from the economic, environmental, and social dimensions [2]. There have been many sustainable development initiatives in different sectors, particularly attempting to improve sustainability by integrating emerging technologies like artificial intelligence, big data, Internet of Things (IoT), and blockchain.

Blockchain is a decentralized information technology based on shared, immutable, distributed ledgers to store transactions for building trust [3]. Although still a rapidly evolving technology, blockchain is a disruptive technology and a digital transformation driver based on a new computing and information flow paradigm [4][5]. Stakeholders have applied blockchain technologies to many domains, from the original application domain of finance, extended to energy, healthcare [3], maritime transport [4], supply chain and logistics [5], industry 4.0 [6], smart cities [7], and others.

The benefits of blockchain include better information traceability, transparency and trust than classical technologies. The blockchain data have better fault tolerance, immutability, tamper prevention, and integrity [8]. Stakeholders have also combined blockchain technology with IoT and artificial intelligence to provide a better service to society [3][9].

Blockchain can be a useful tool to address issues related to sustainability. A critical use case for blockchain-based applications is "*tracking potential social and environmental conditions that might pose environmental, health, and safety concerns*" [5]. Traceability has been recognized to have a significant role in contributing to SDGs, particularly SDG 8, 9, and 12 [10]. Studies show that blockchain technology enables technology to improve traceability for sustainability in manufacturing and supply chains [4][8]. Blockchain also makes it possible to verify sustainability with quantifiable and meaningful sustainability indicators [8]. Thus, there is increasing research on applying blockchain technology to address sustainability.

## II. THE CHALLENGE PROBLEM

Blockchain can improve sustainability but can potentially bring negative impacts to sustainable development [8]. As blockchain is an emerging technology, studies address the adoption barriers from various aspects [4][5][8][9]. However, to our knowledge, there is a lack of a body of knowledge to systematically summarize the benefits, challenges, enablers, and barriers of using blockchain technologies to achieve SDGs.

In an ongoing study, we want to investigate "*what is the relationship between blockchain and sustainability? How does blockchain technology contribute to sustainable development, e.g., related to SDGs?*" In detail, we want to answer the following research questions (RQs):

- RQ1: What are the positive and negative impacts on SDGs using blockchain technologies in various applications?

Blockchain technology's ability to enable secure data exchange in a distributed manner without intermediaries profoundly impacts supply chains in terms of organizations, supply chain relationships, and transactions. Moreover, blockchain technologies facilitate supply chain visibility and contribute to product traceability, authenticity, and legitimacy [10]. Blockchain technology's traceability advantage is also regarded as a critical enabler for smart cities with efficient supply chains, dynamic business interaction, and increased consumer satisfaction and trust [7]. Although these blockchain-based applications can lead to economic benefits to the stakeholders, few studies have investigated the added SDGs values using blockchain in different applications. Blockchain can also bring negative impacts to SDGs. For example, the Proof-of-Work consensus algorithms are regarded by many people as not environmentally friendly because they require many calculations and use a lot of energy and carbon footprint [8]. Summarizing the negative impact of blockchain on SDGs may also bring insights on blockchain technology development.

- RQ2: What are the key enablers and barriers to adopting blockchain technologies in practice to satisfy SDG goals?

As with any emerging technologies, blockchain technology's adoption faces not only technological barriers due to its immaturity but also other barriers related to organizational, social, legal, and regulatory aspects [4][5][8][9]. One example adoption barrier is the unwillingness of partners to share valuable and critical information to protect competitiveness. A systematic overview of the relevant adoption barriers and risks can guide technology adopters and business decision-makers through the journey of sustainability development, e.g., tackling the challenges with the right set of policies and regulations. Thus, we aim to identify social, cultural, and ethical enablers and barriers to use blockchain technologies to satisfy SDG goals.

- RQ3: What are the research gaps to facilitate sustainability using blockchain technologies?

The results of RQ1 and RQ2 will illustrate the state-of-the-art of using blockchain technologies for sustainability development. It is then necessary to identify research gaps from the results to motivate further development of blockchain technologies.

III. TERTIARY STUDY

We have made a pilot search in the Scopus database using the keywords "blockchain" AND "sustainability OR sustainable." By limiting the results to reviews and journal publications alone, we have got 34 review articles. 21 out of the 34 papers were published in 2020. The high increase of the recent publications indicates the rapid development and increasing interest in this topic. However, the high number of existing reviews also suggests that it may be more beneficial to start with consolidating existing reviews' results than to do a new systematic literature review. Therefore, we decided to perform a tertiary study to answer our research questions.

A tertiary study or tertiary review is a review of secondary studies (e.g., systematic literature reviews) related to the same research questions [13]. In tertiary research, researchers usually use thematic analysis methods to extract information from existing reviews to answer research questions. We have done a pilot analysis of 18 included literature review papers using thematic analysis[1]. The pilot analysis has shown several exciting results:

- We found that several reviews have classified selected publications from social, economic, and environmental dimensions of sustainability (triple-bottom-line) [4][12]. However, no study has systematically summarized the relationship between blockchain and sustainability in terms of SDGs.

- For RQ1, we found some positive connections between blockchain and SDGs, e.g., traceability and responsible sourcing, for fulfilling SDG 8, 9, 12, and 14 (in particular SDG targets 14.1 to 14.5) [8]. Blockchain-based applications for financial inclusion also contributes to SDG 1 and 8. Studies also pointed out the negative impact of blockchain's energy intensiveness on SDG 7 and 13 [14].

- For RQ2, results showed a wide range of barriers, e.g., energy intensiveness and the "*lack of government and industrial policies and willingness to guide and encourage sustainable and safe practices*" [4].

- For RQ3, we found that 11 of the 18 included literature reviews are on supply chains (in general or specific supply chains like agriculture, food, or agri-food) and supply chain management. As blockchain can benefit other fields such as healthcare, mobility, e-voting, and energy [3][7], it would be interesting to investigate if the supply chains' findings apply to other fields. So far, we have not found reviews addressing software engineering aspects. Empirical software engineering research and evaluation of blockchain-based applications will thus be a valuable future research topic.

Our future work is to continue the thematic data analysis by following the tertiary study guidelines in [13]. Based on identified challenges, barriers, and research gaps from the results of our research questions, we will perform follow-up studies to investigate them in depth.

ACKNOWLEDGMENT

This work is jointly supported by the National Key Research and Development Program of China (No. 2019YFE0105500) and the Research Council of Norway (No. 309494 and 274816).

---

[1] The search query and analysis from the pilot search is available at: https://github.com/SINTEF-SE/P4C/blob/main/Scopus202012analysis.pdf